\documentclass{appolb}

\usepackage{epsfig}
\usepackage{amssymb}

\def\oo{\'o}
\def\pb1{pb$^{-1}$}
\def\as{\alpha_s}
\def\q2{Q^2}
\def\ele{e^+e^-}
\def\pp{p\bar p}
\def\z0{Z^0}
\def\qq{q\bar q}
\def\colab#1{#1 Coll.}
\def\kt{k_T}
\def\g2{GeV$^2$}
\def\etjb{E^{\rm jet}_{T,{\rm B}}}
\def\oalphas2{{\cal O}(\alpha\as^2)}
\def\asz{\as(\mz)}
\def\asmz#1#2#3#4#5#6{\asz = #1\pm #2\ {\rm (stat.)}\ ^{+#4}_{-#3}\ {\rm (exp.)}\ ^{+#6}_{-#5}\ {\rm (th.)}}
\def\mz{M_Z}
\def\etjet{E_T^{\rm jet}}
\def\xo{x_{\gamma}^{\rm obs}}
\def\cost{\vert\cos\theta^*\vert}
\def\etajet{\eta^{\rm jet}}
\def\set{d\sigma/d\etjet}
\def\sxo{d\sigma/d\xo}
\def\seta{d\sigma/d\etajet}
\def\m3j{M^{\rm 3j}}

\def\Journal#1#2#3#4{{#1} {#2} (#3) #4}

\def\PLB{{\em Phys. Lett.}  {\bf B}}

\def\EPC{{\em Eur. Phys. Jour.} {\bf C}}

\begin{document}

\title{Hadronic final states and QCD studies at HERA\footnote{Talk
    given at ``Physics at LHC'', 3-8 July 2006, Cracow, Poland.}
}
\author{C. Glasman\thanks{Ram\oo n y Cajal Fellow.}\\
 (on behalf of the ZEUS and H1 Collaborations)
\address{Universidad Aut\oo noma de Madrid}}
\maketitle
\begin{abstract}
Results on QCD studies from the H1 and ZEUS Collaborations at the $ep$
collider HERA are presented and their impact on LHC physics discussed.
\end{abstract}

%\vspace{-0.5cm}
\section{Introduction}
HERA is the only electron(positron)-proton collider in the world. It
has been running since 1992, with a major upgrade of the accelerator
during 2001-2002. The multipurpose H1 and ZEUS detectors have been
collecting data steadily during these years. Pre-upgrade integrated
luminosities of $\sim 100$~\pb1\ per experiment have already allowed
the most accurate determination of the proton structure up to
date. Both collaborations have also made important contributions in
making tests and precision measurements of QCD and the electroweak
sector of the Standard Model, search for new particles and new
interactions, study of jet production and jet substructure, study of
the structure of the photon, heavy flavour production, and
investigation of diffraction and the structure of the pomeron.

This report deals with the tests of QCD on hadronic final states,
concentrating mainly on the results from jet production. Multijet QCD
production will be the main background for searches of Higgs,
supersymmetric particles and new physics in general. Also, the
luminosity of the colliding particles (partons from the protons) at
LHC will be governed by QCD and so precise determinations of the
protons parton distribution functions (PDFs) and the strong coupling
constant, $\as$, will be crucial for any cross section measurement. In
this sense, $ep$ collisions at HERA provide a suitable enviroment to
do precision studies of QCD in hadronic-induced reactions (as opposed
to $\ele$ annihilations at LEP) in a cleaner set-up than in $\pp$
collisions at Tevatron.

The main sources of jets at HERA are neutral current (NC) deep
inelastic scattering (DIS), with $\q2\gg\Lambda^2_{\rm QCD}$, where
$\q2$ is the square of the momentum transfer, and photoproduction (PHP),
with $\q2\approx 0$. Up to leading order (LO) in $\as$, jet production
in NC DIS proceeds via the quark-parton model ($Vq\rightarrow q$, where
$V=\gamma^*$ or $\z0$), boson-gluon fusion ($Vg\rightarrow \qq$) and
QCD-Compton ($Vq\rightarrow qg$) processes. The jet production
cross section is given in perturbative QCD (pQCD) by the convolution
of the proton PDFs and the partonic cross section. Only two kinematic
variables are needed to characterise these processes; they can be
taken as $\q2$ and Bjorken $x$. The inelasticity variable $y$, which
gives the energy transfer in the proton rest frame, is also useful. In
PHP, a quasi-real photon emitted by the electron beam interacts with a
parton from the proton to produce two jets in the final state. In LO
QCD, there are two processes which contribute to the jet
photoproduction cross section: the resolved process, in which the
photon interacts through its partonic content, and the direct process,
in which the photon interacts as a point-like particle. The jet
production cross section is given by the convolution of the flux of
photons in the electron, the parton densities in the proton and photon
and the partonic cross section. Thus, measurements of jet cross
sections in NC DIS and PHP can be used to perform tests of pQCD, help
to constrain the gluon density in the proton and determine $\as$.

%Figure 2
\begin{figure}[th]
\setlength{\unitlength}{1.0cm}
\begin{picture} (10.0,6.5)
\put (-0.5,0.5){\epsfig{figure=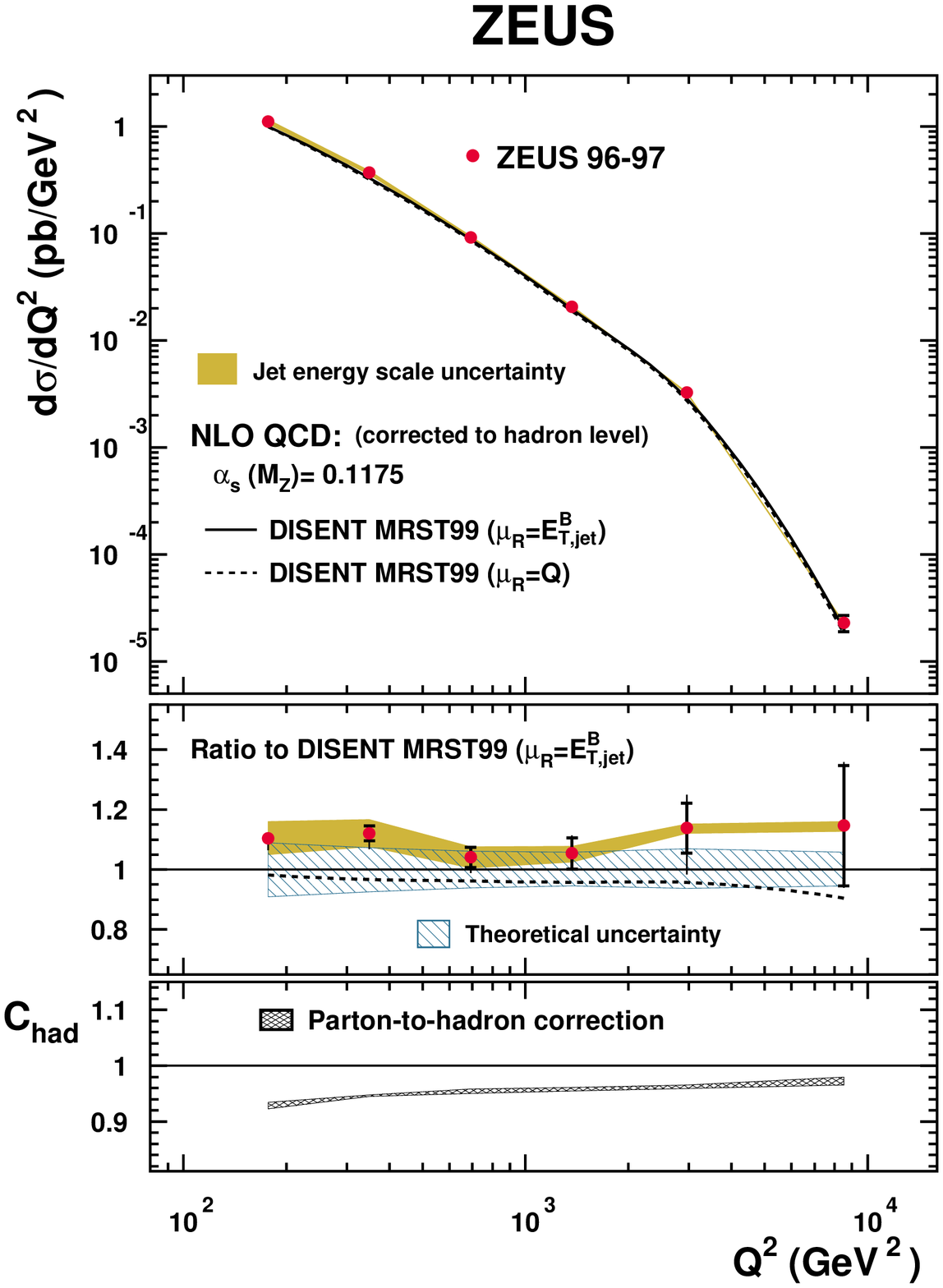,width=5.0cm}}
\put (3.9,0.5){\epsfig{figure=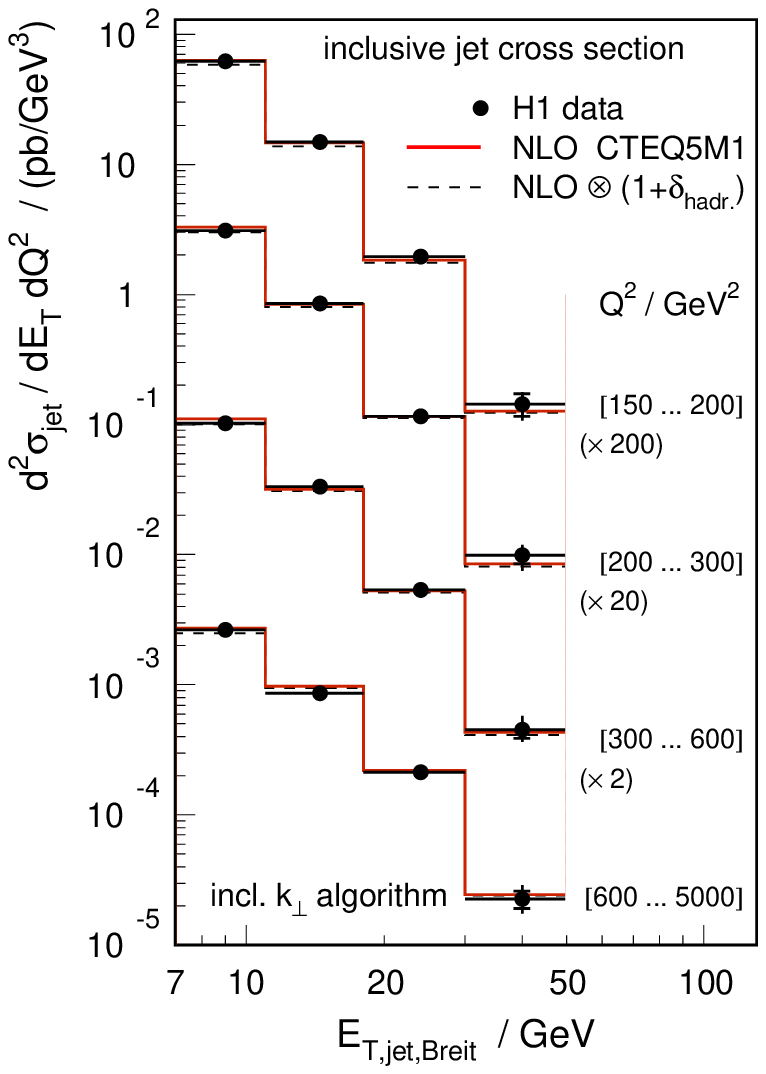,width=4.2cm}}
\put (7.5,0.5){\epsfig{figure=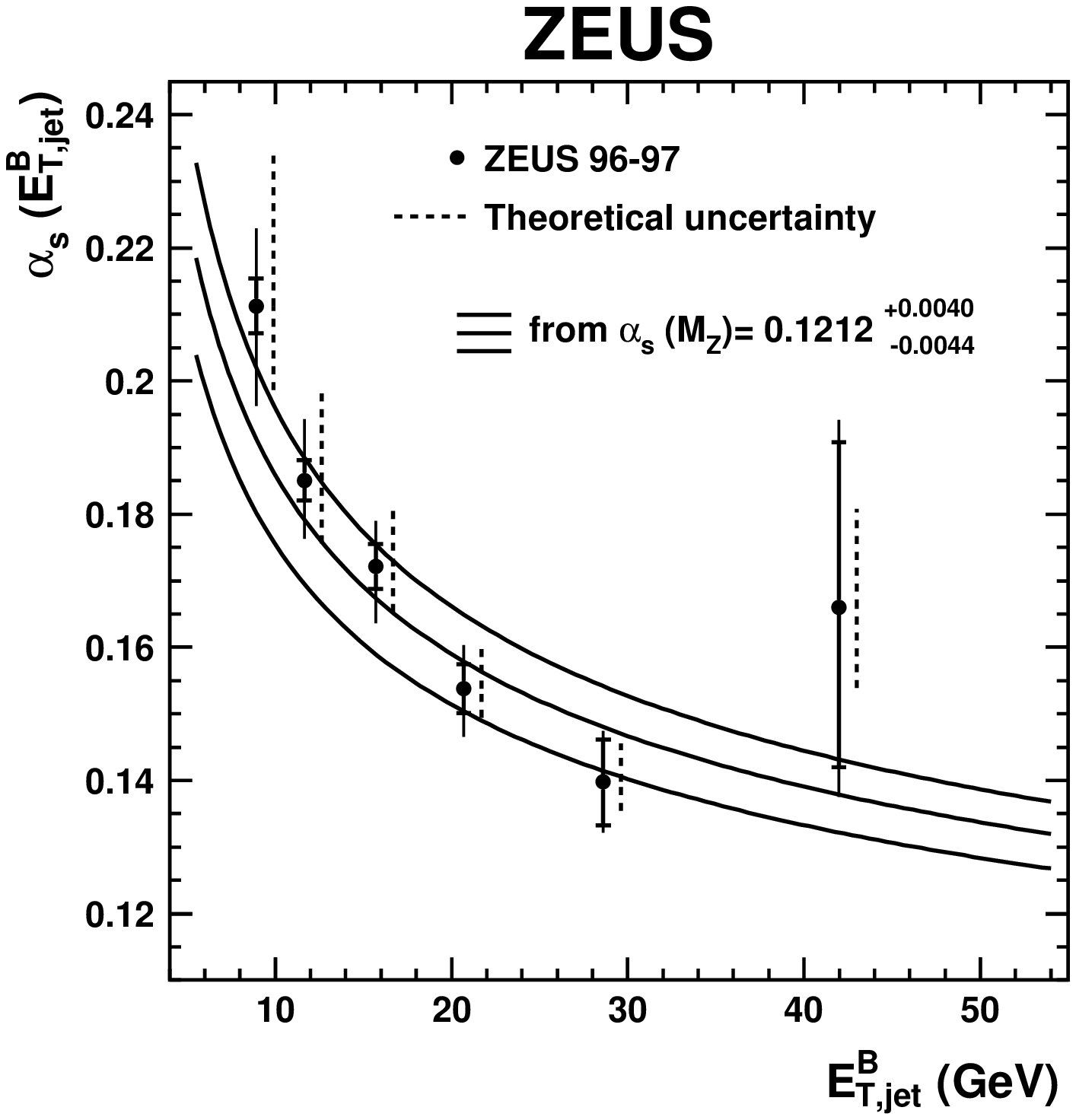,width=6.5cm}}
\put (2.0,0.0){\bf\small (a)}
\put (6.0,0.0){\bf\small (b)}
\put (10.5,0.0){\bf\small (c)}
\end{picture}
\caption{(a) Inclusive-jet cross section in NC DIS as a function of
  $\q2$~\protect\cite{pl:b547:164}; (b) inclusive-jet cross section in
  NC DIS as a function of $\etjb$ in different regions of
  $\q2$~\protect\cite{epj:c19:289}; (c) energy-scale dependence of
  $\as$~\protect\cite{pl:b547:164}.
  \label{fig2}}
\end{figure}

%Figure 3
\begin{figure}[th]
\setlength{\unitlength}{1.0cm}
\begin{picture} (10.0,6.0)
\put (-0.5,0.5){\epsfig{figure=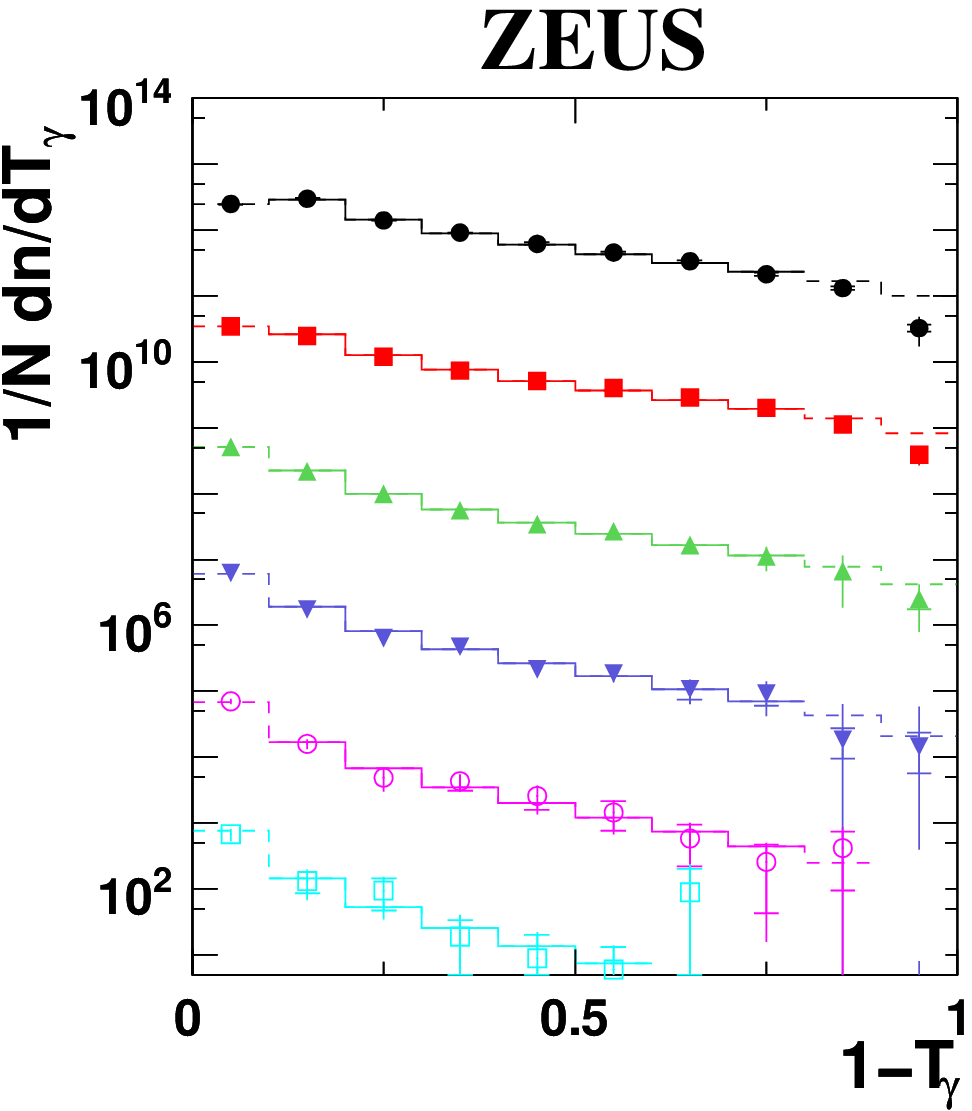,width=5.0cm}}
\put (5.9,0.2){\epsfig{figure=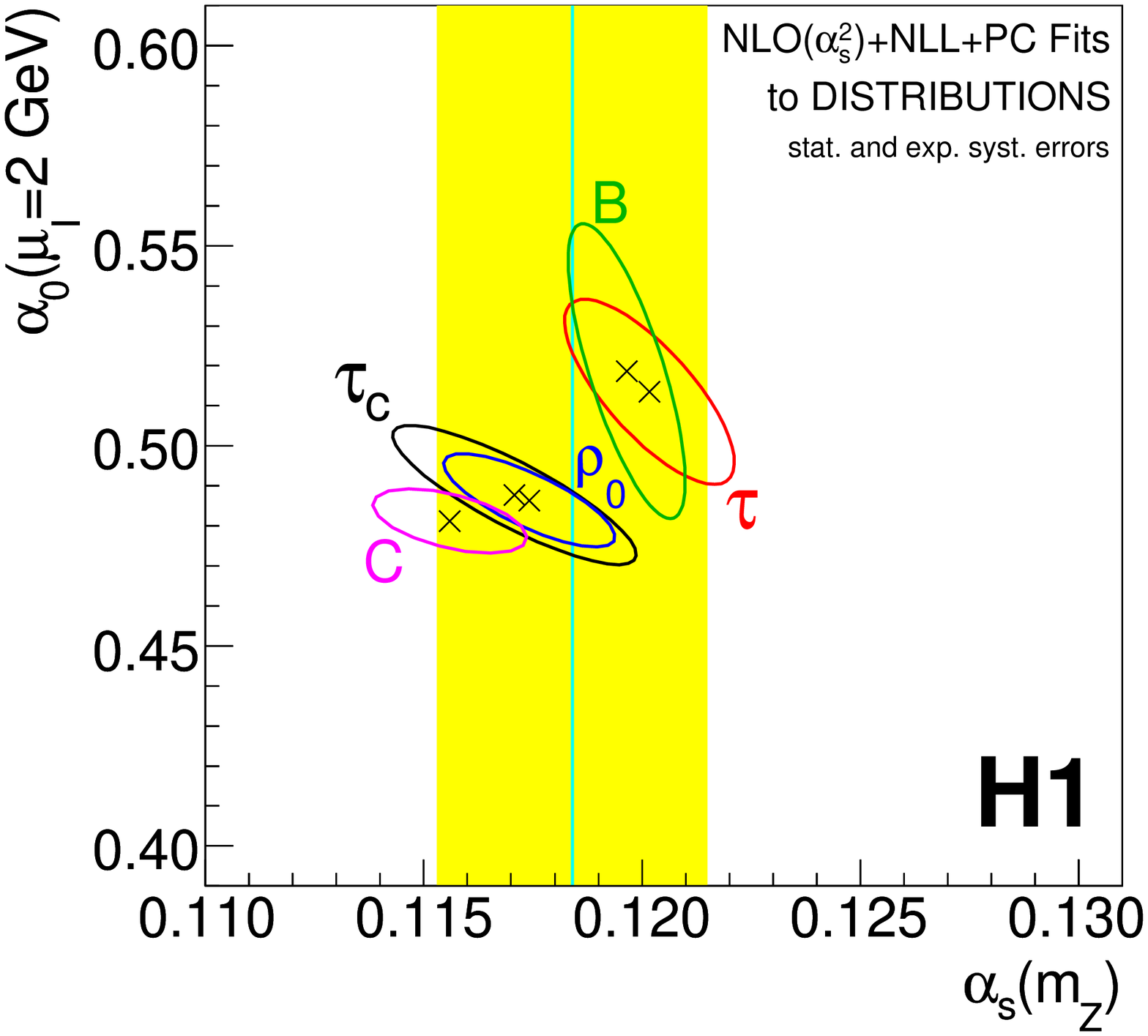,width=6.2cm}}
\put (2.0,0.0){\bf\small (a)}
\put (9.0,0.0){\bf\small (b)}
\end{picture}
\caption{(a) Thrust distribution in different $\q2$
  regions~\protect\cite{desy-06-042}; (b) $\bar\alpha_0$ and $\asz$
  values extracted from event-shape
  distributions~\protect\cite{epj:c46:343}.
  \label{fig3}}
\end{figure}

%Figure 4
\begin{figure}[th]
\setlength{\unitlength}{1.0cm}
\begin{picture} (10.0,4.5)
\put (7.0,0.0){\epsfig{figure=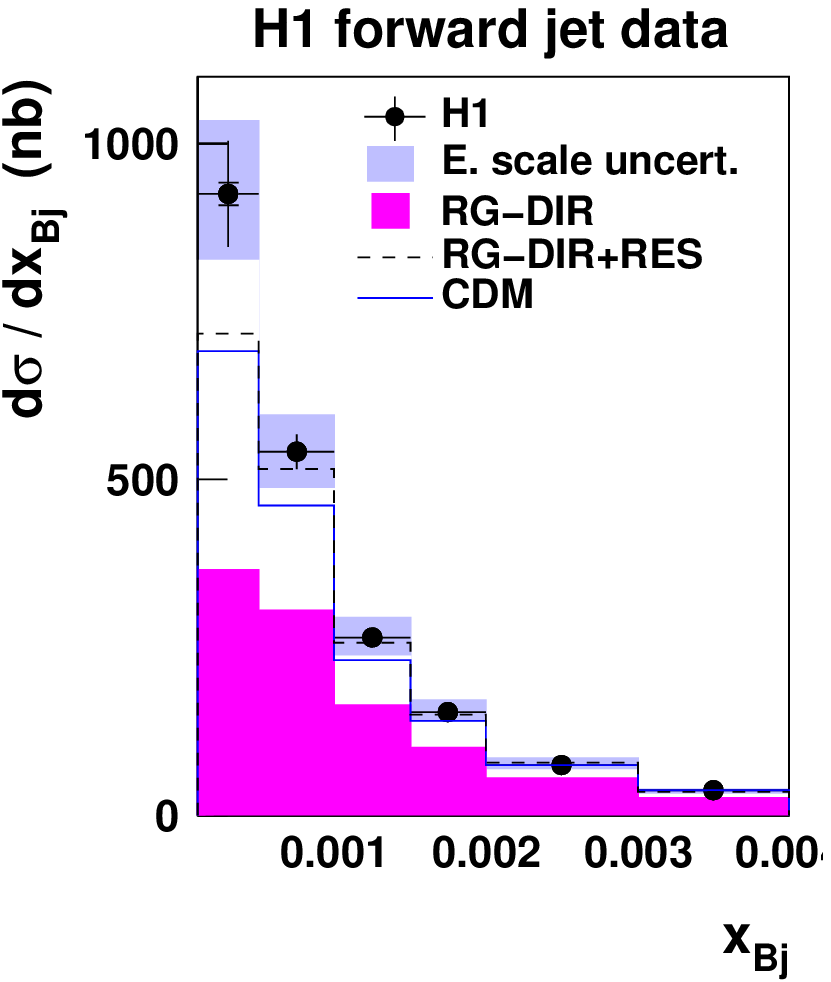,width=4.0cm}}
\put (5.9,0.7){\epsfig{figure=white.eps,width=2.0cm,height=4.0cm}}
\put (4.125,0.0){\epsfig{figure=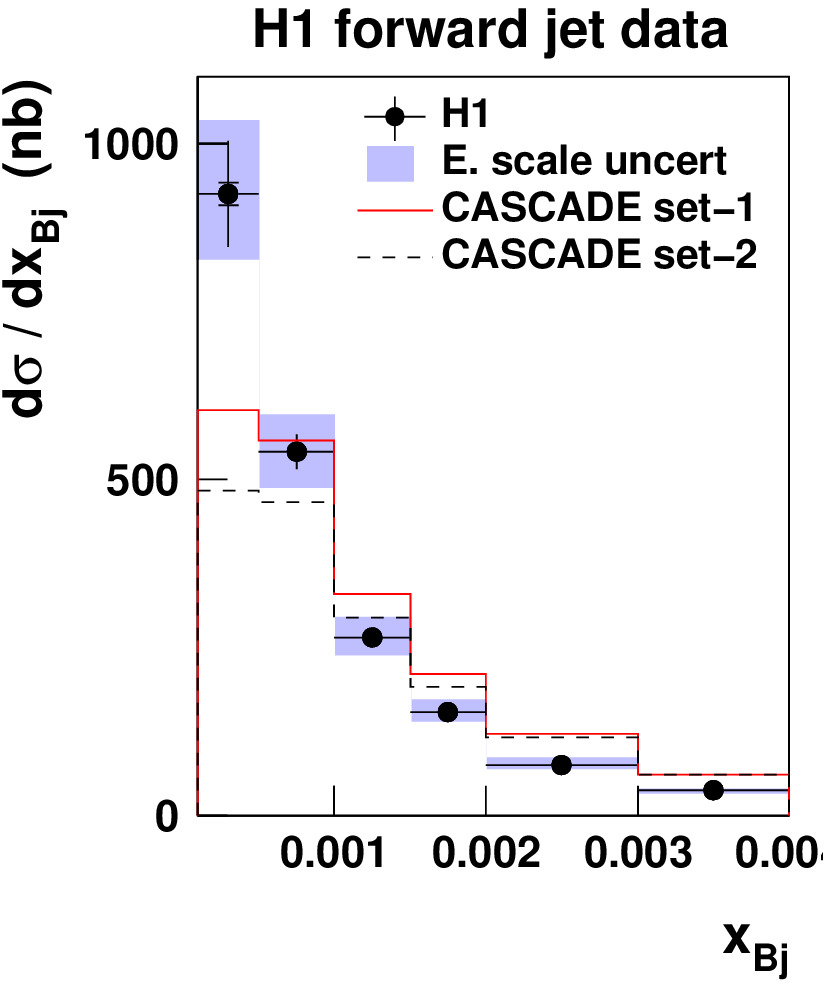,width=4.0cm}}
\put (3.0,0.7){\epsfig{figure=white.eps,width=2.0cm,height=4.0cm}}
\put (1.2,0.0){\epsfig{figure=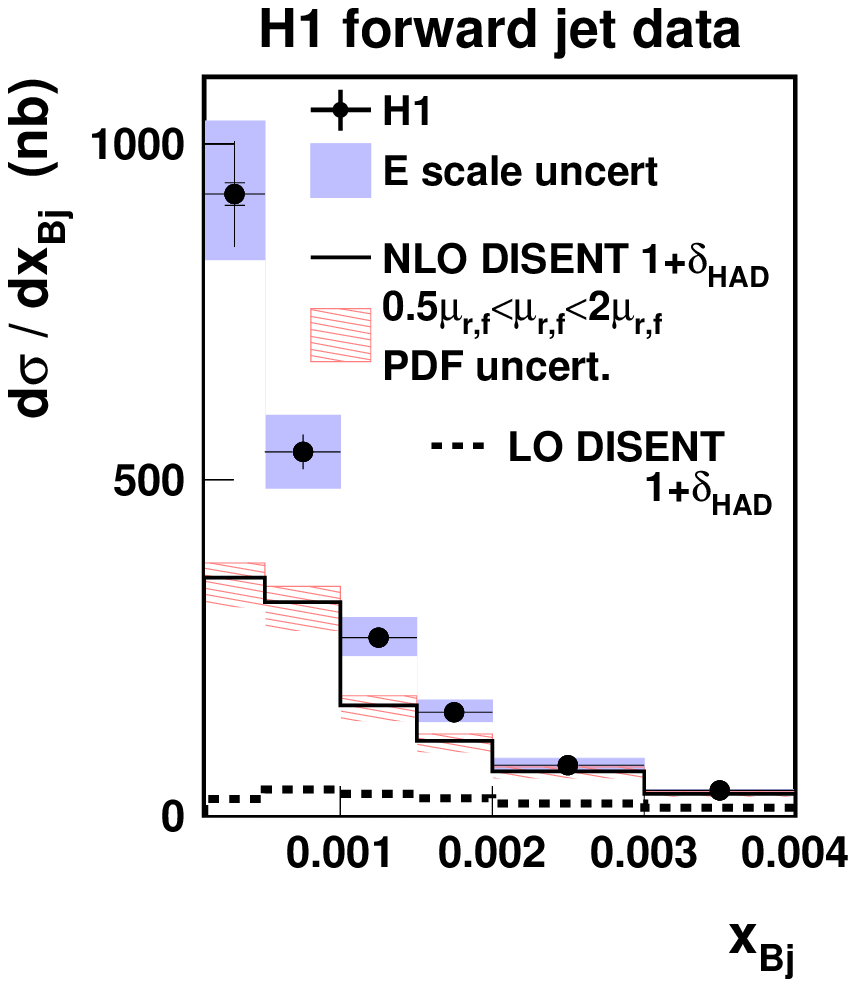,width=4.175cm}}
\put (3.5,0.0){\bf\small (a)}
\put (6.0,0.0){\bf\small (b)}
\put (9.0,0.0){\bf\small (c)}
\end{picture}
\caption{Forward-jet cross section as a function of Borken
  $x$~\protect\cite{epj:c46:27} compared to (a) LO and NLO 
  calculations; (b) {\sc Casacade} MC predictions; and (c) CDM and
  {\sc Rapgap} MC predictions.
  \label{fig4}}
\end{figure}

%Figure 5
\begin{figure}[th]
\setlength{\unitlength}{1.0cm}
\begin{picture} (10.0,5.0)
\put (4.0,0.0){\epsfig{figure=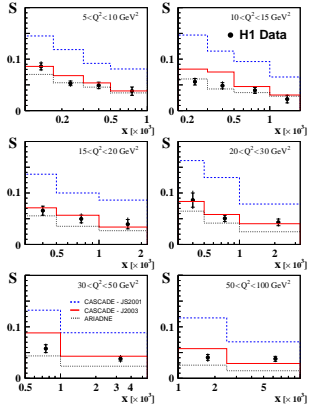,width=4.0cm}}
\end{picture}
\caption{Azimuthal correlation as a function of $x$ for different
  $\q2$ regions~\protect\cite{epj:c33:477} compared to the 
  CDM and {\sc Casacade} MC predictions.
  \label{fig5}}
\end{figure}

%Figure 6
\begin{figure}[th]
\setlength{\unitlength}{1.0cm}
\begin{picture} (10.0,5.0)
\put (1.0,0.1){\epsfig{figure=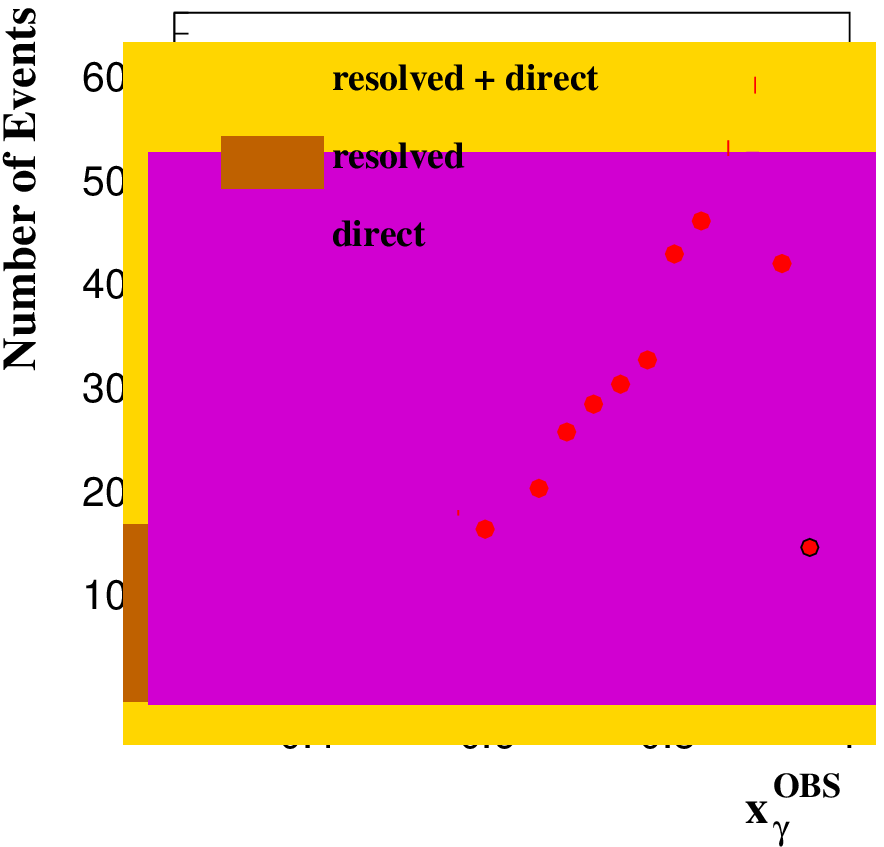,width=5.0cm}}
\put (1.0,0.1){\epsfig{figure=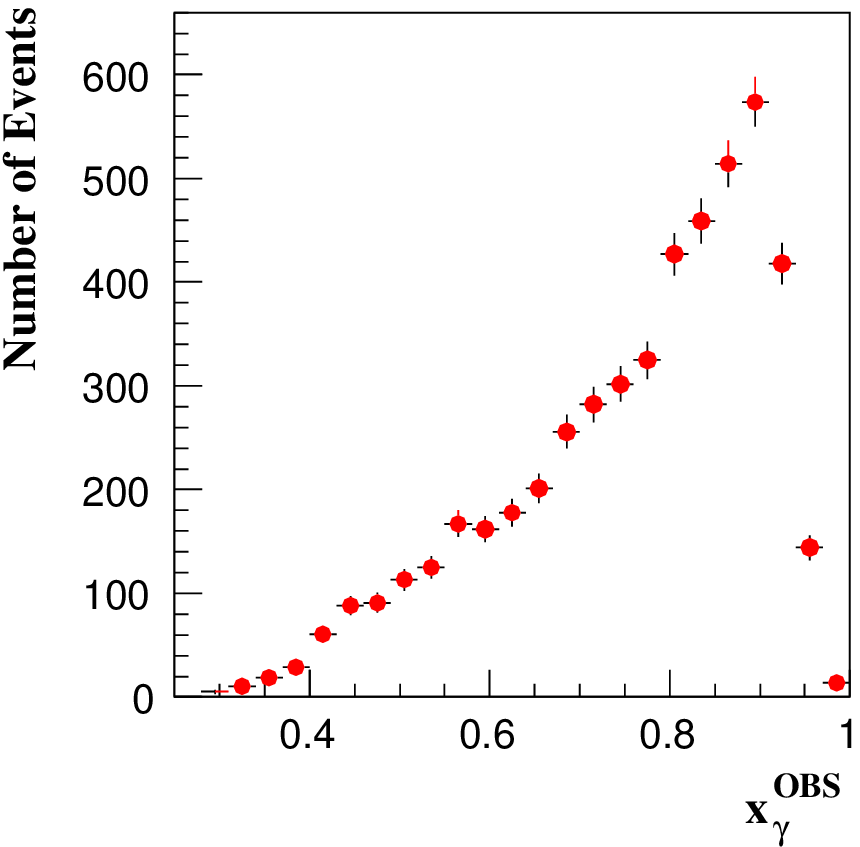,width=5.0cm}}
\put (2.3,-1.3){\epsfig{figure=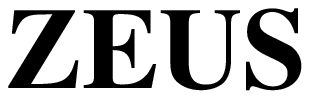,width=8.0cm}}
\put (6.4,0.2){\epsfig{figure=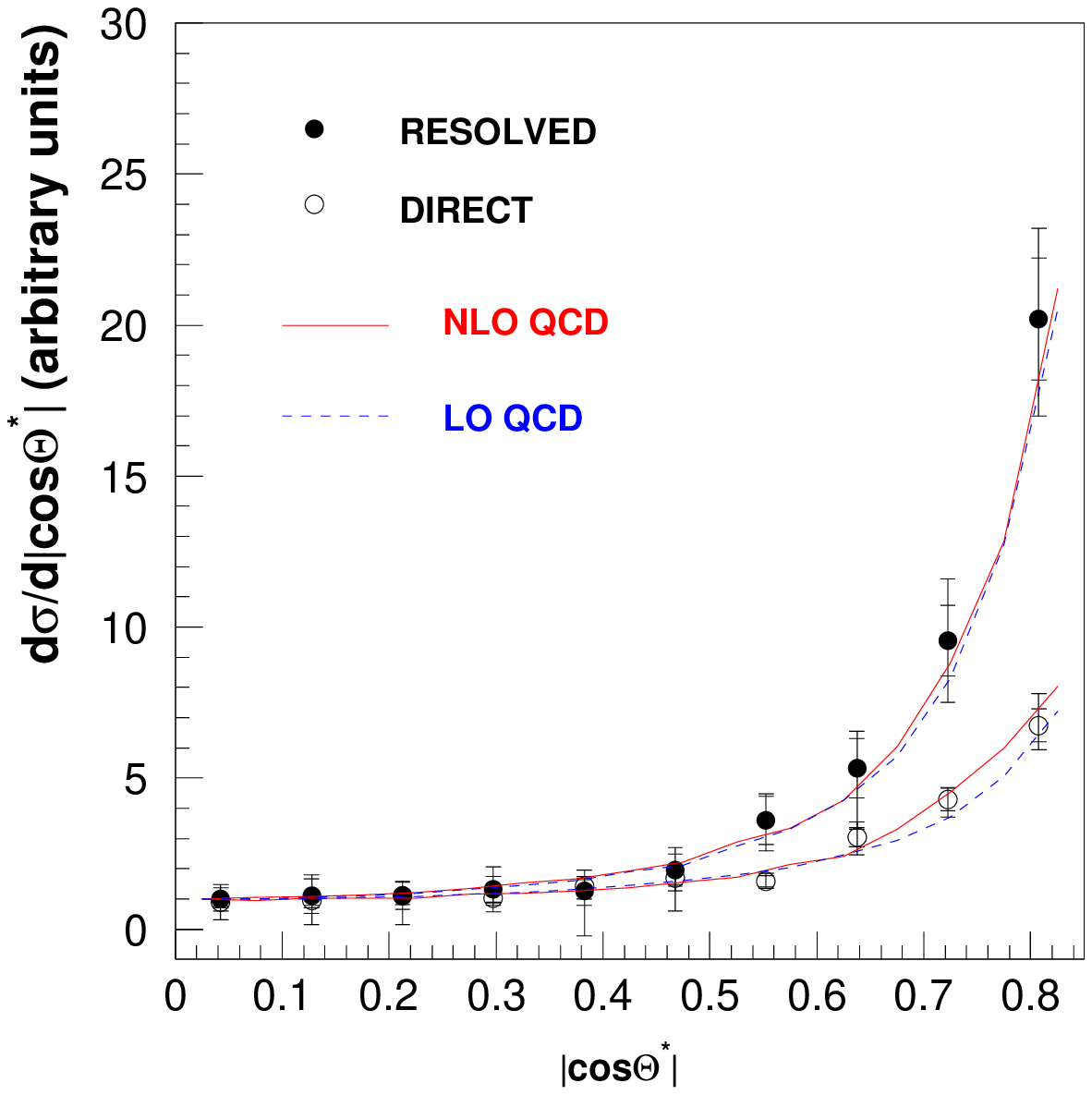,width=5.2cm}}
\put (3.0,0.0){\bf\small (a)}
\put (9.0,0.0){\bf\small (b)}
\end{picture}
\caption{(a) $\xo$ distribution for data and resolved, direct and
  resolved+direct MC events~\protect\cite{pl:b384:401}; (b) Dijet
  cross section as a function of $\cost$ in PHP for 
  $\xo\lessgtr 0.75$~\protect\cite{pl:b384:401}.
  \label{fig6}}
\end{figure}

%Figure 7
\begin{figure}[th]
\setlength{\unitlength}{1.0cm}
\begin{picture} (10.0,5.0)
\put (-0.5,0.2){\epsfig{figure=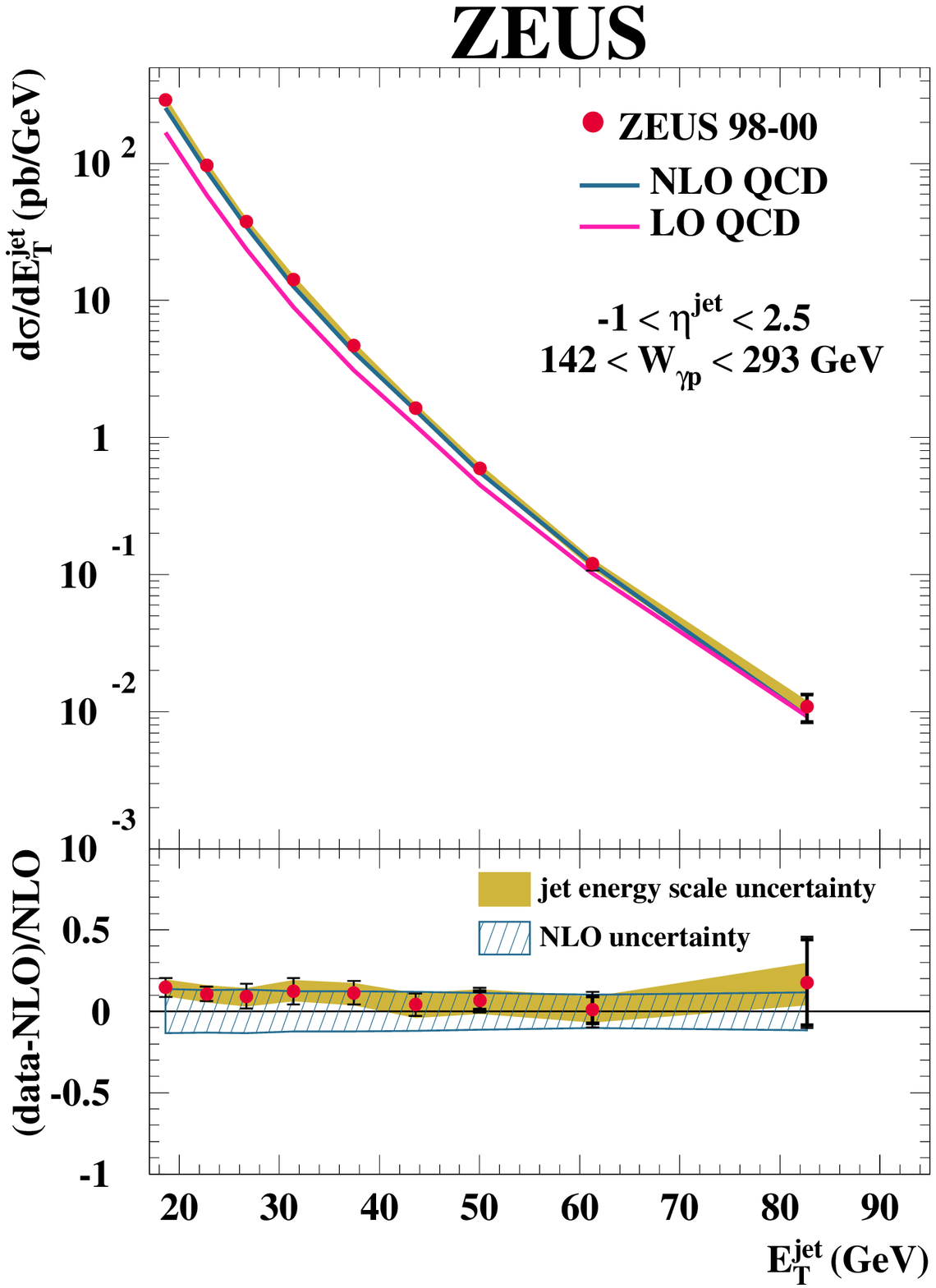,width=5.5cm}}
\put (3.6,0.2){\epsfig{figure=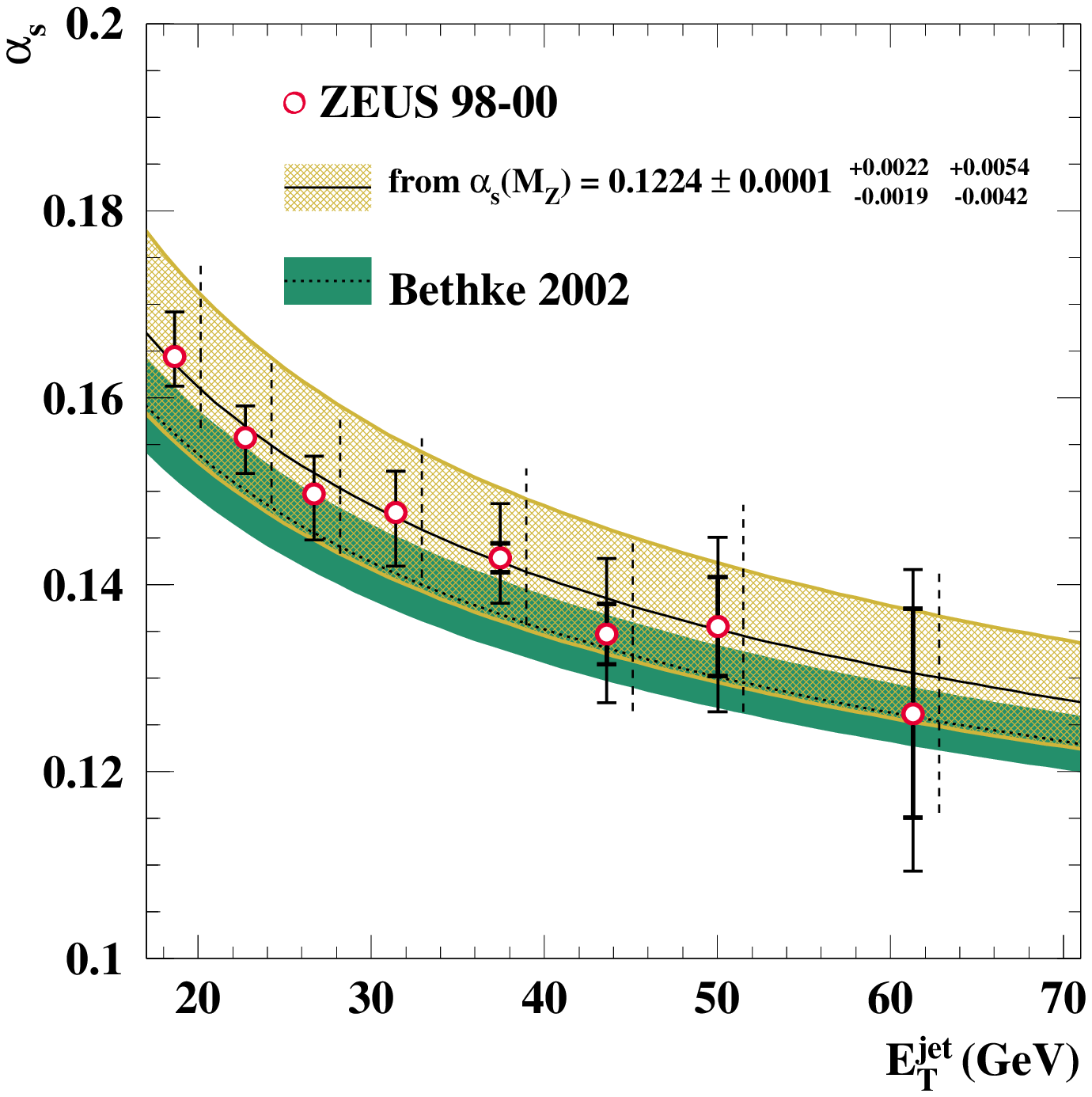,width=5.8cm}}
\put (8.5,0.2){\epsfig{figure=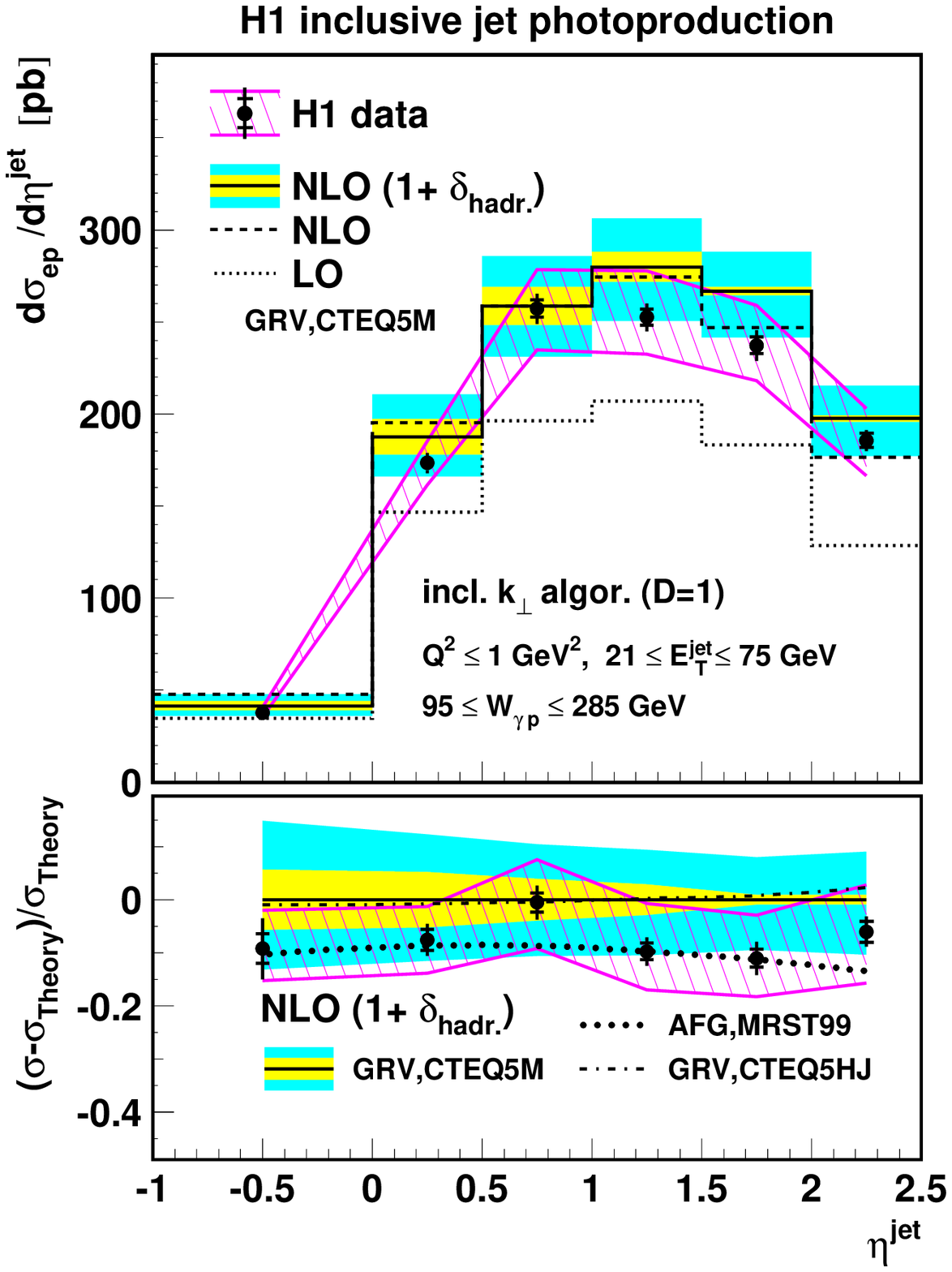,width=4.0cm}}
\put (2.0,0.0){\bf\small (a)}
\put (6.0,0.0){\bf\small (b)}
\put (10.5,0.0){\bf\small (c)}
\end{picture}
\caption{(a) Inclusive-jet cross section as a function of 
  $\etjet$~\protect\cite{pl:b560:7}; (b) $\as$ as a function of
  $\etjet$~\protect\cite{pl:b560:7}; (c) inclusive-jet cross sections
  as a function of $\etajet$~\protect\cite{epj:c29:497}.
  \label{fig7}}
\end{figure}

%Figure 8
\begin{figure}[th]
\setlength{\unitlength}{1.0cm}
\begin{picture} (10.0,5.0)
\put (0.5,0.1){\epsfig{figure=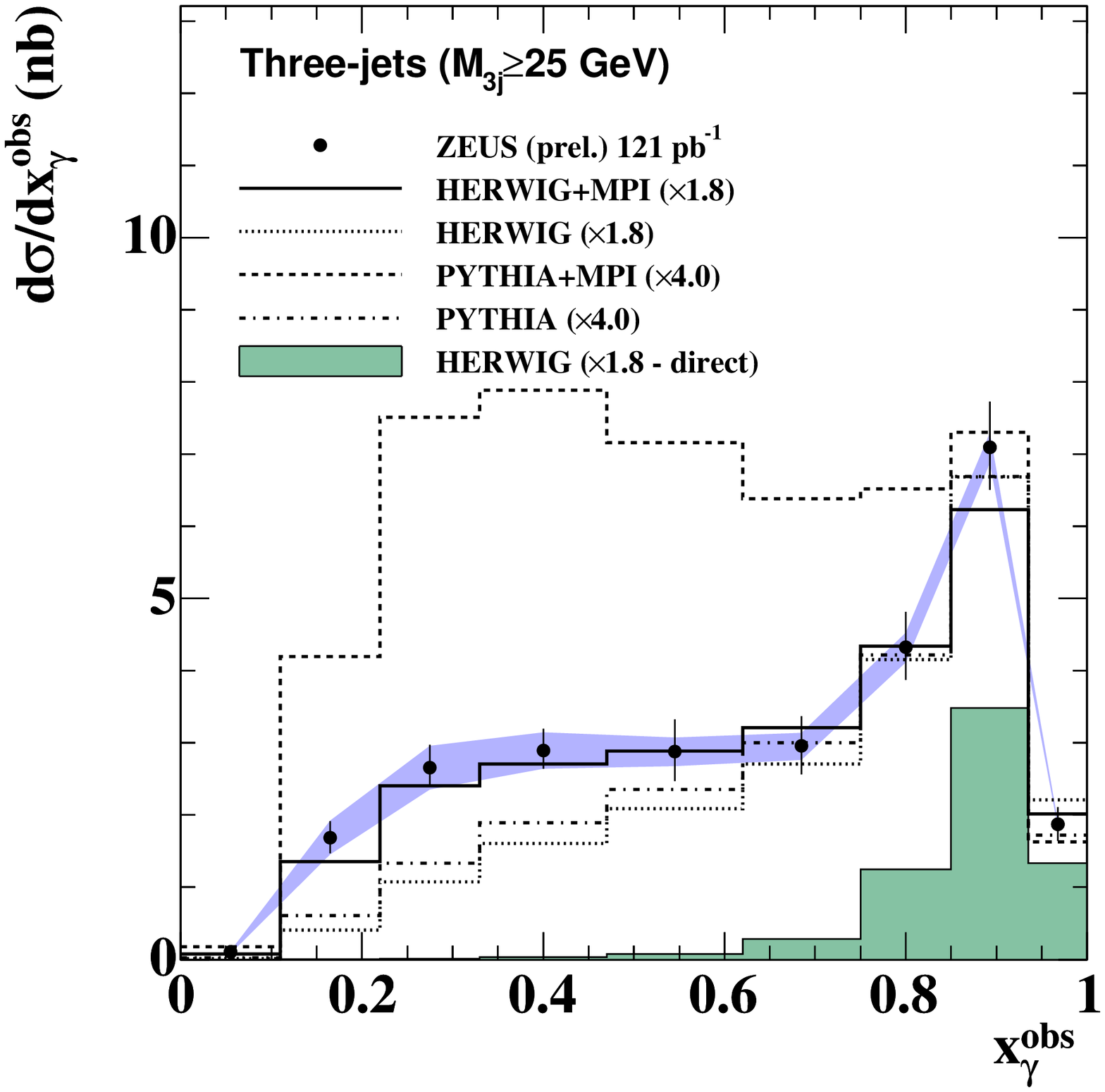,width=5.0cm}}
\put (6.4,0.1){\epsfig{figure=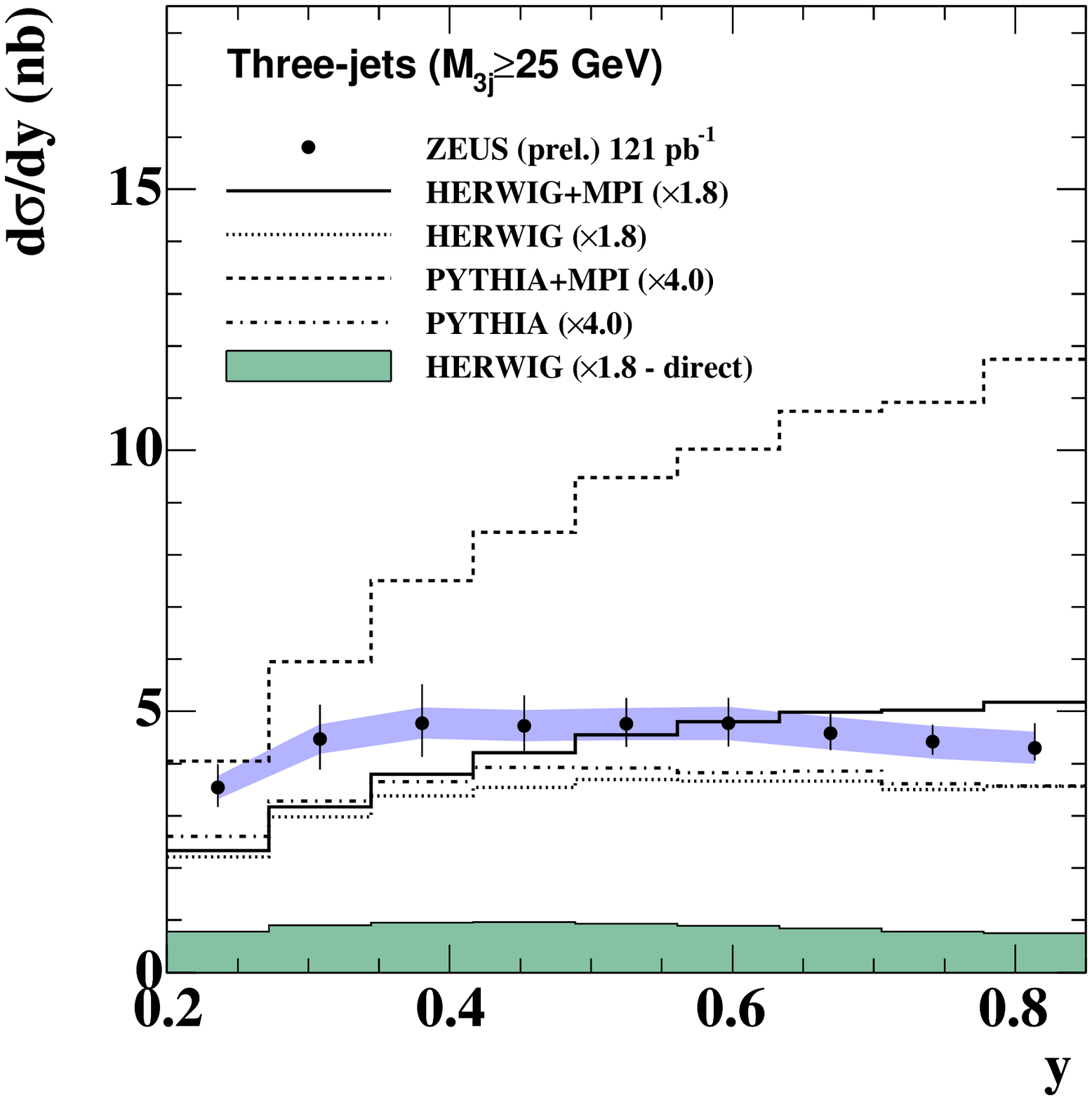,width=5.0cm}}
\put (2.3,-1.3){\epsfig{figure=zeus.eps,width=8.0cm}}
\put (3.0,0.0){\bf\small (a)}
\put (9.0,0.0){\bf\small (b)}
\end{picture}
\caption{Three-jet cross sections in PHP for $\m3j>25$ GeV as
  functions of (a) $\xo$ and (b)
  $y$~\protect\cite{zeus-prel-06-009}. For comparison, the predictions
  of {\sc Pythia} and {\sc Herwig} with and without MPIs are included.
  \label{fig8}}
\end{figure}

%Figure 9
\begin{figure}[th]
\setlength{\unitlength}{1.0cm}
\begin{picture} (10.0,6.0)
\put (0.0,0.1){\epsfig{figure=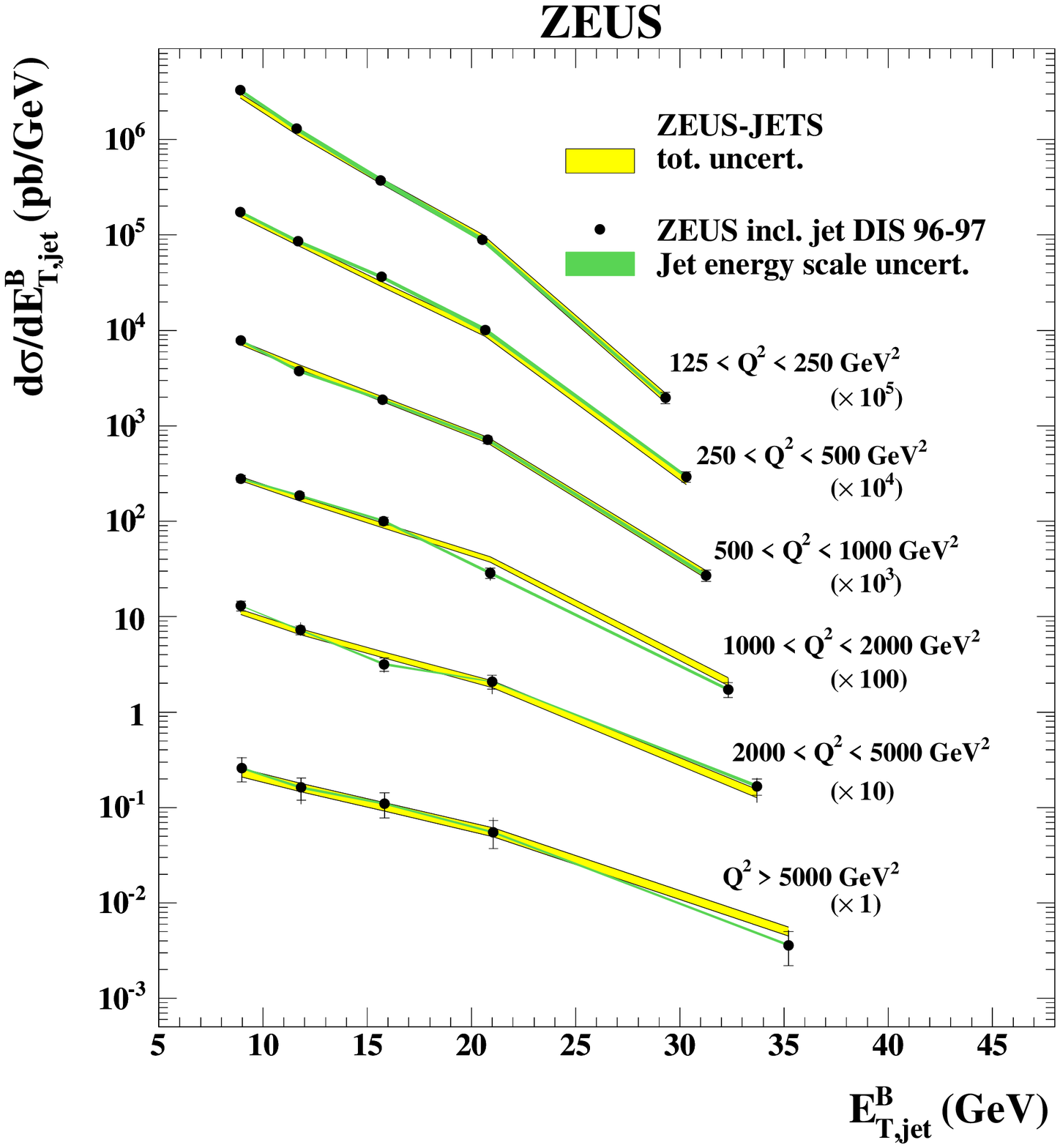,width=5.3cm}}
\put (6.4,0.2){\epsfig{figure=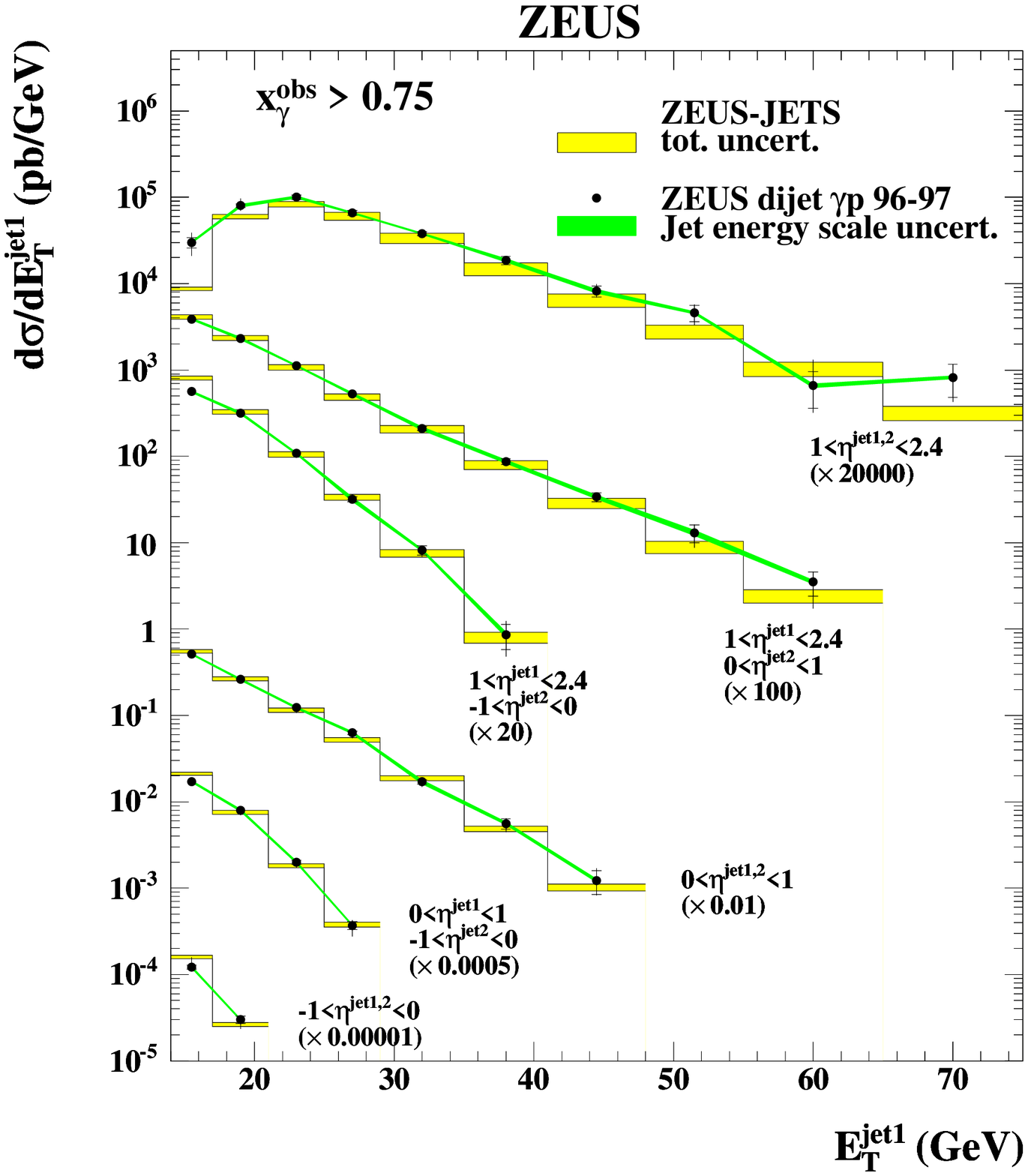,width=5.3cm}}
\put (3.0,0.0){\bf\small (a)}
\put (9.0,0.0){\bf\small (b)}
\end{picture}
\caption{(a) Inclusive-jet cross section in NC DIS as a function of
  $\etjb$ in different regions of $\q2$~\protect\cite{pl:b547:164};
  (b) Dijet cross section in PHP as a function of $\etjet$ in
  different regions of $\etajet$ for $\xo> 0.75$~\protect\cite{pl:b384:401}.
  Both measurements are compared to NLO calculations based on
  ZEUS-jets proton PDFs~\protect\cite{epj:c42:1}.
  \label{fig9}}
\end{figure}

%Figure 10
\begin{figure}[th]
\setlength{\unitlength}{1.0cm}
\begin{picture} (10.0,6.0)
\put (0.5,-3.0){\epsfig{figure=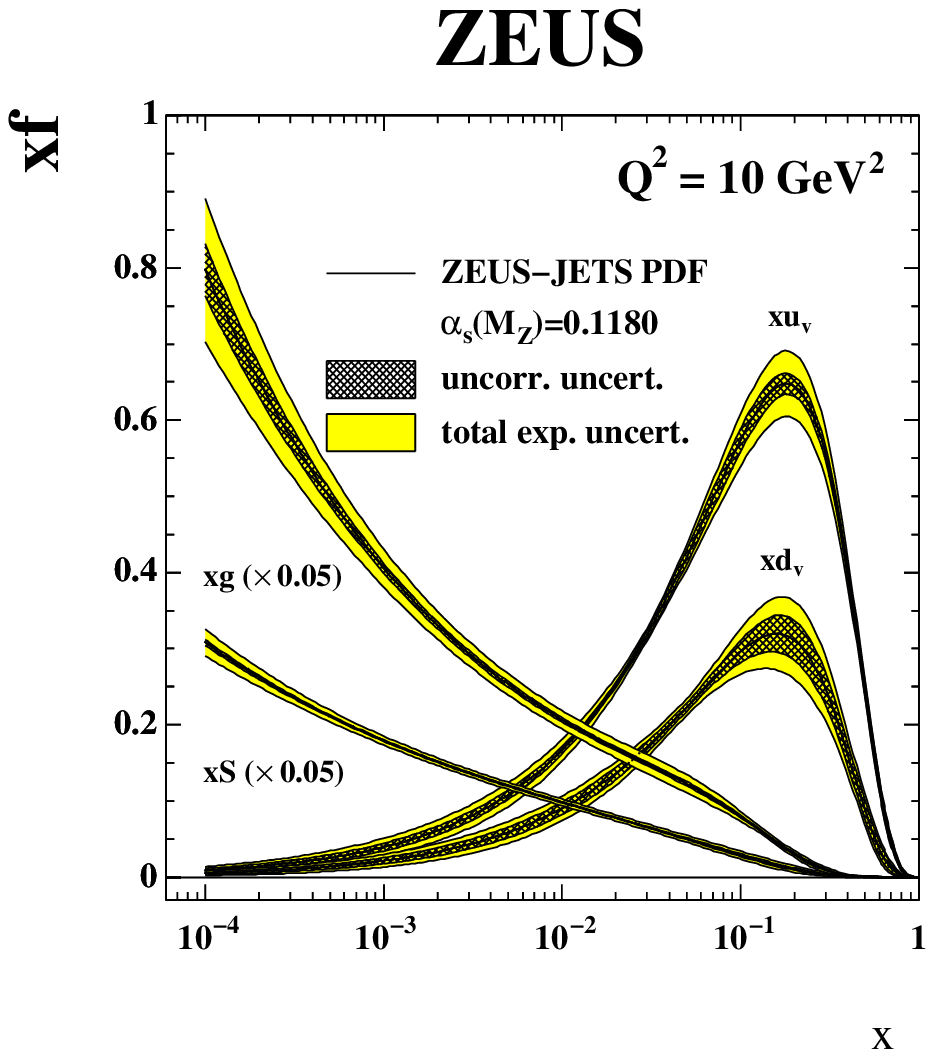,width=7.7cm}}
\put (6.4,0.2){\epsfig{figure=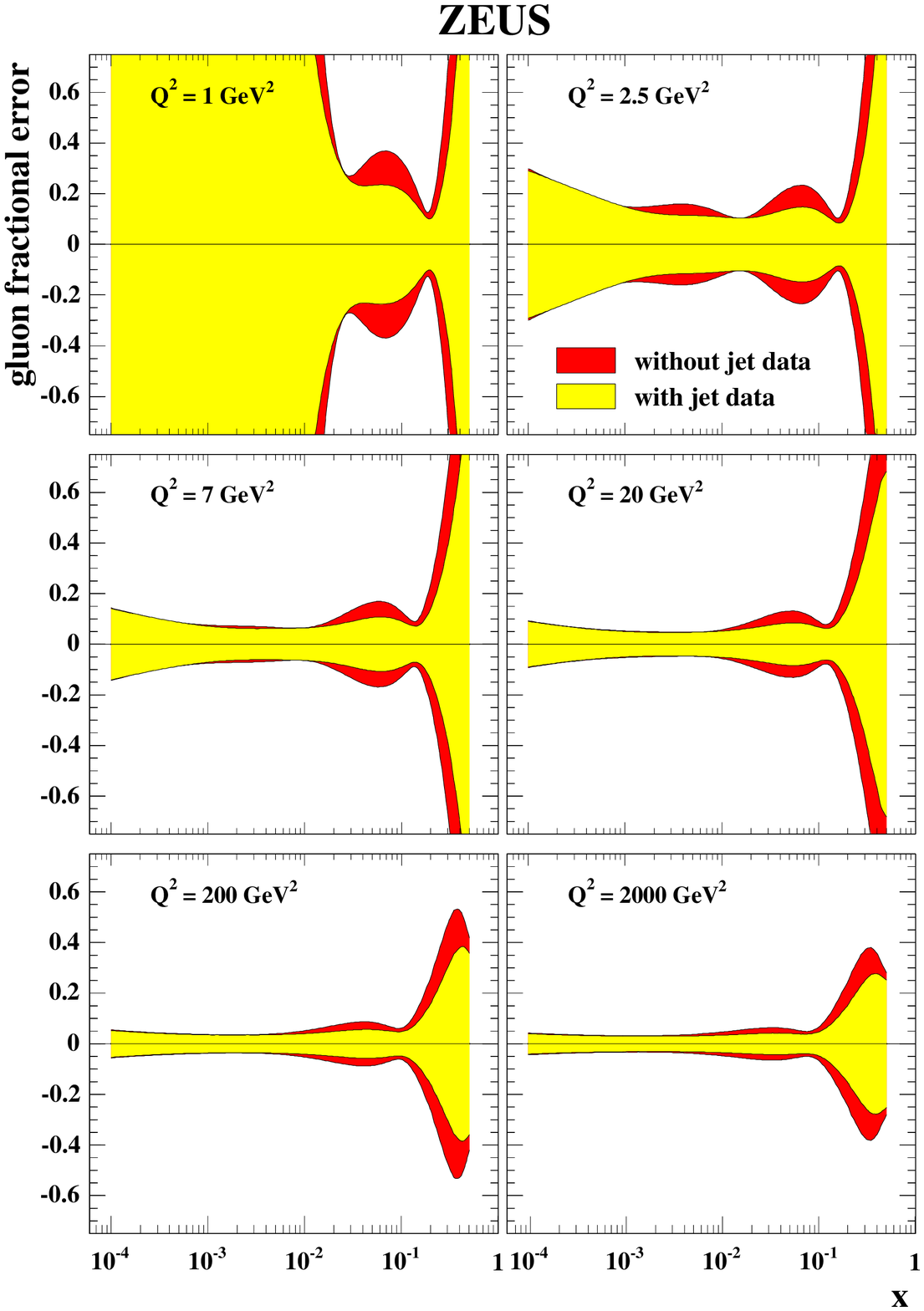,width=4.0cm}}
\put (3.0,0.0){\bf\small (a)}
\put (9.0,0.0){\bf\small (b)}
\end{picture}
\caption{(a) Valence, sea and gluon distributions from the ZEUS-jets
  proton PDF parametrisations~\protect\cite{epj:c42:1};
  (b) uncertainties of the ZEUS-jets
  proton PDF parametrisations~\protect\cite{epj:c42:1}.
  \label{fig10}}
\end{figure}

%Figure 11
\begin{figure}[th]
\setlength{\unitlength}{1.0cm}
\begin{picture} (10.0,7.0)
\put (0.5,0.0){\epsfig{figure=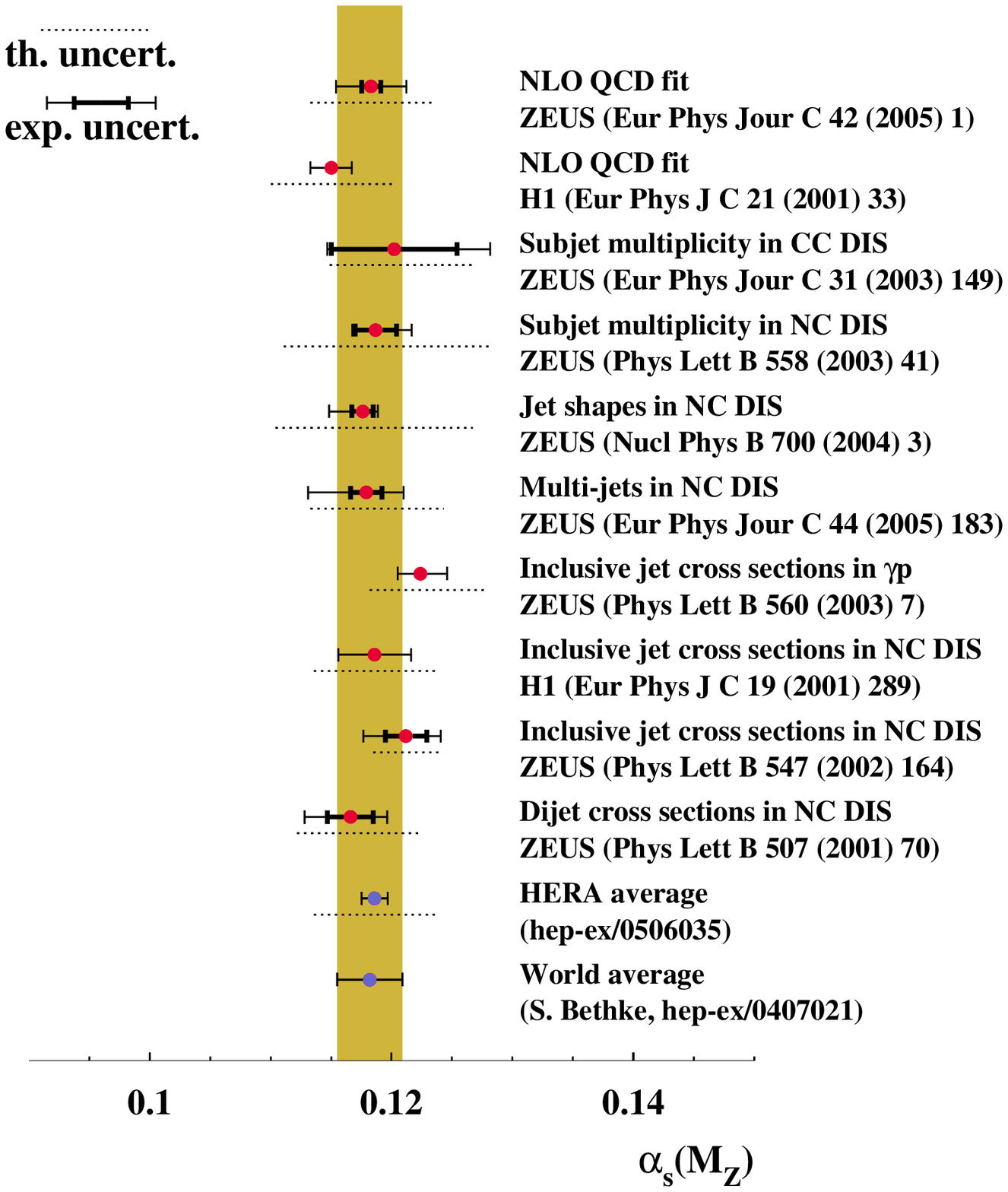,width=7.7cm}}
\put (8.4,4.2){\epsfig{figure=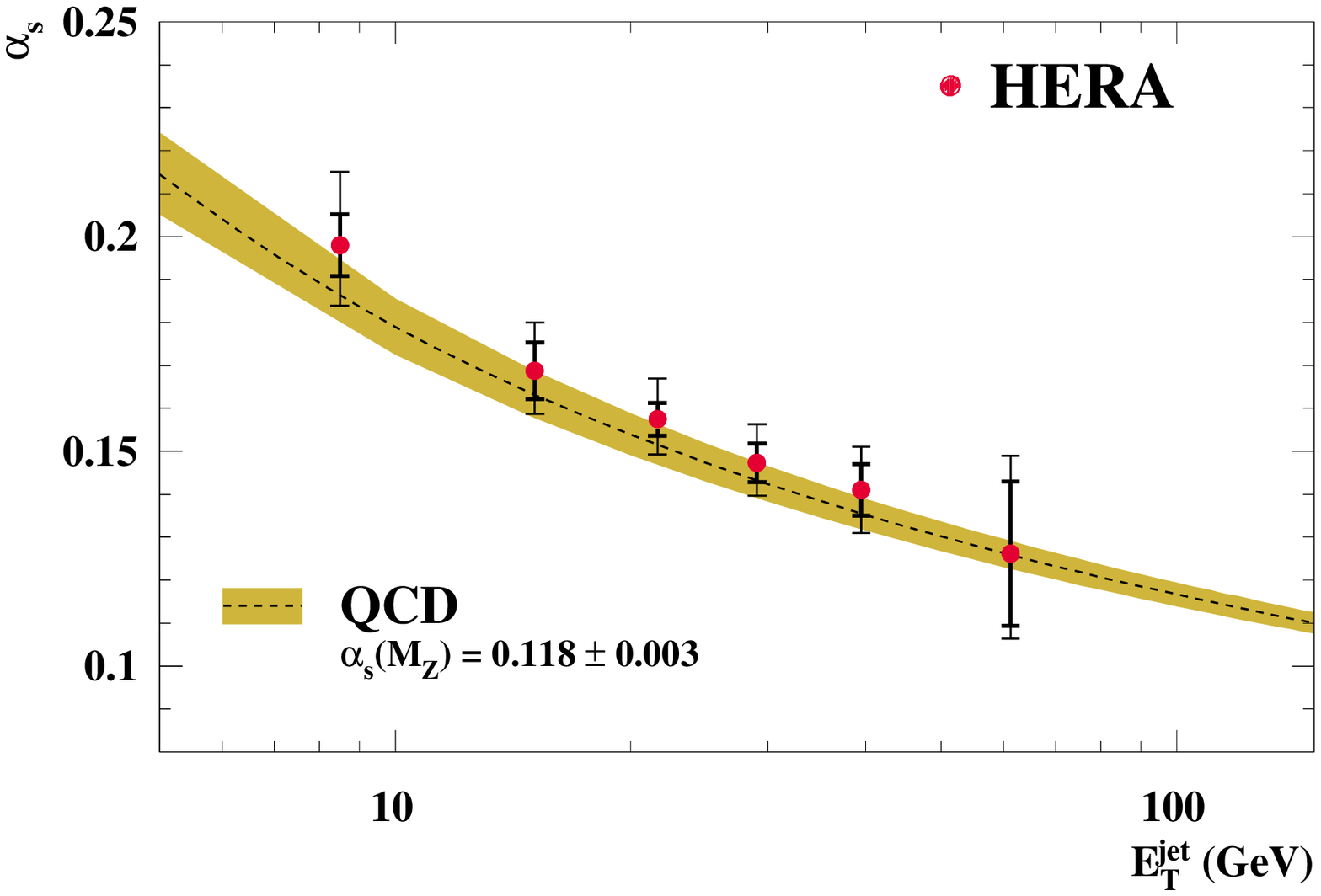,width=4.0cm}}
\put (8.5,0.2){\epsfig{figure=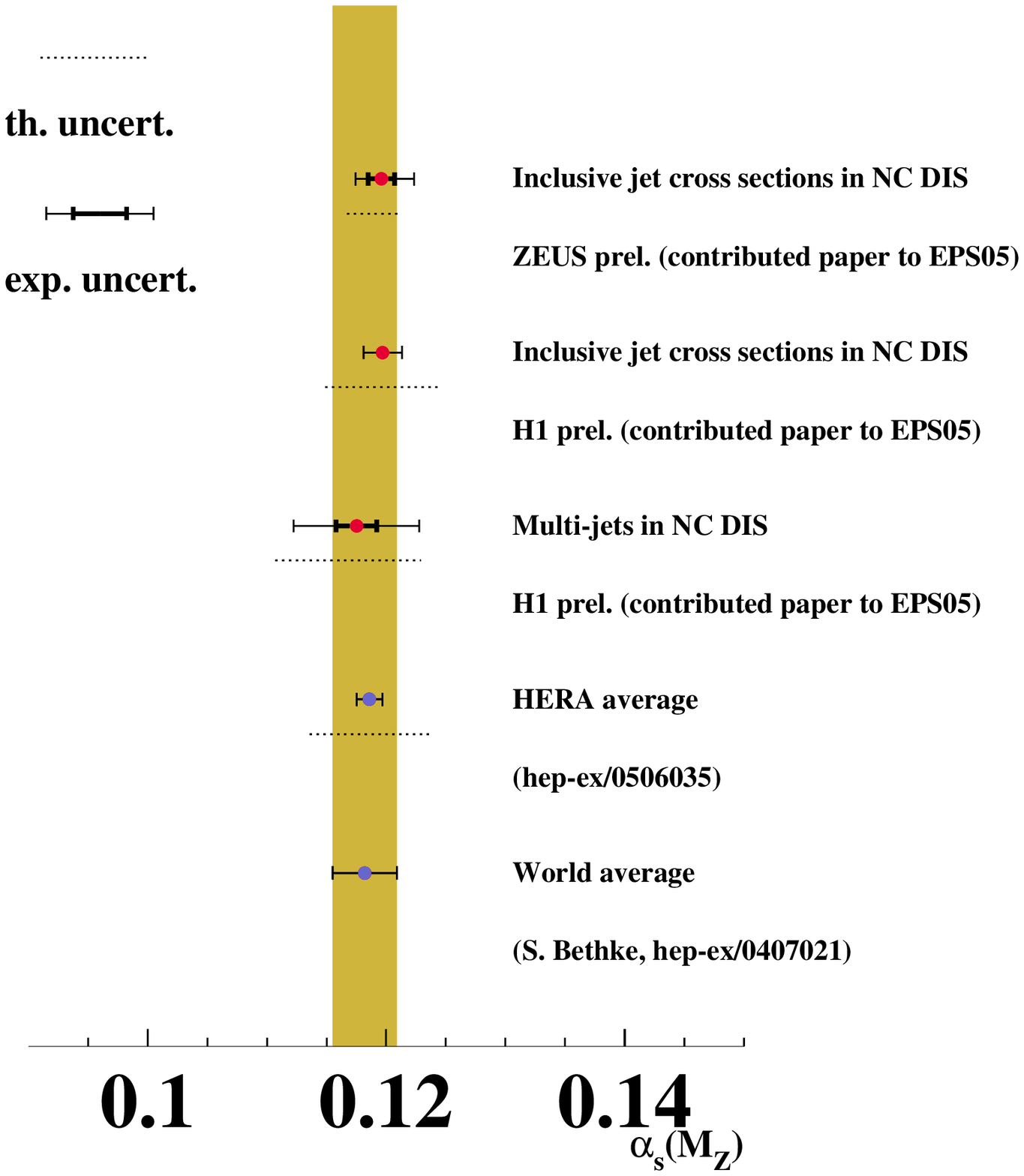,width=4.7cm}}
\put (3.0,0.0){\bf\small (a)}
\put (10.0,4.6){\bf\small (b)}
\put (10.0,0.0){\bf\small (c)}
\end{picture}
\caption{(a,b) Summary of $\asz$ measurements at HERA compared with the
  world and HERA averages~\protect\cite{herave};
  (b) combined measurements of $\as(\etjet)$ from
  HERA~\protect\cite{herave}.
  \label{fig11}}
\end{figure}

%\vspace{-0.5cm}
\section{Jet search in $ep$ collisions}
Jets in hadronic collisions are not as easily identified as in $\ele$
annihilations since the initial-state colliding partons carry only a
fraction of the energy of the parent hadrons and are accompanied by
several soft hadrons not correlated with the hard interaction. The
spectator partons give rise to the so-called remnant jets
(hadronisation of spectator partons), underlying event (soft
interactions between the spectator partons) and multiparton
interactions (hard interactions between the spectator partons).

The collisions in hadronic-type reactions do not occur in the
centre-of-mass frame; the initial-state partonic system is boosted
along the beam axis by an amount which is different for every
event. To treat on equal footing all possible final-state hadronic
systems, variables invariant under longitudinal boosts, such as
transverse energy ($E_T$), differences in pseudorapidity ($\eta$) and
azimuth ($\phi$), are best suited to reconstruct jets in this type of
reactions. Also, the use of the transverse energy helps to identify the
signal for a hard scattering and to disentangle the products of the
hard interaction from the beam remnants. These features apply to both
NC DIS and PHP processes at HERA, though in NC DIS there is a further
complication due to the fact that the exchanged boson is virtual and
carries $p_T$. This effect is compensated for by selecting a frame in
which the virtual boson and the proton collide head-on. This is the
so-called Breit frame.

At HERA, the $\kt$ cluster algorithm in the longitudinal inclusive
mode has been proven to be the best algorithm to reconstruct jets in
these hadronic-type reactions. This algorithm has been used by both
the H1 and ZEUS collaborations since many years for making presicion
tests of pQCD. The advantages of this algorithm can be summarisedas
follows: ($i$) it allows a transparent translation of the experimental
set-up to the theoretical calculations (it avoids the ambiguities
related to overlapping and merging of jets present in cone
algorithms); ($ii$) pQCD calculations using this algorithm are finite
to all orders (as opposed to the cone algorithm, for which the
calculations diverge for orders higher than the next-to-leading); and
($iii$) it presents the smallest hadronisation corrections compared to
other algorithms.
% (see Fig.~\ref{fig1}). 
This algorithm has been used
to reconstruct the jets in data and theoretical predictions for the
measurements presented in the next sections.

%\vspace{-0.5cm}
\section{QCD studies in neutral current deep inelastic scattering}

%\vspace{-0.1cm}
\subsection{Inclusive-jet cross sections and $\as$}
Inclusive-jet cross sections have been measured in NC DIS at large
$\q2$ ($\q2>125$~\g2~\cite{pl:b547:164} and
$\q2>150$~\g2~\cite{epj:c19:289}) as functions of $\q2$, the jet
transverse energy in the Breit frame ($\etjb$) and for $\etjb$ in
different regions of $\q2$ (see Fig.~\ref{fig2}). The measured cross
sections as functions of $\q2$ and $\etjb$ display a steep fall-off of
several orders of magnitude within the measured range; the data show a
harder spectrum in $\etjb$ as $\q2$ increases. The measurements have
been compared to next-to-leading-order (NLO) QCD calculations. The
predictions give a good description of the data over a wide range of
$\q2$ and $\etjb$. This proves the validity of the description of the
dynamics of inclusive-jet production by pQCD at $\oalphas2$. The LO
prediction for the jet cross section in the Breit frame is directly
proportional to $\as$. Values of $\asz$ have been extracted from the
measurements:

\centerline {$\asmz{0.1212}{0.0017}{0.0031}{0.0023}{0.0027}{0.0028}$
  ($\q2>500$~\g2),}

\centerline {$\asz=0.1186\pm 0.0030\ {\rm (exp.)}\pm 0.0051\ {\rm
    (th.)}$ ($\q2>150$~\g2),}
The small experimental ($2.9\%$ and $2.5\%$, respectively) and
theoretical ($2.3\%$ and $4.3\%$, respectively) uncertainties make
these determinations the most precise at HERA and comparable to those
values extracted from more inclusive measurements. The energy-scale
dependence of $\as$ has been tested from the measurements of the
inclusive-jet cross section as a function of $\etjb$. The results,
shown in Fig.~\ref{fig2}(c), are in very good agreement with the
running of $\as$ as predicted by pQCD over a wide range in $\etjb$.

%\vspace{-0.5cm}
\subsection{Event shapes and the hadronisation process}
The hadronic final-state in NC DIS has also been used to study the
hadronisation process, which is a non-perturbative effect. Recent
developments in the model of power corrections permit the
understanding of these effects from first principles. The model of
Dokshitzer \etal\ is characterised by an effective universal coupling,
$\bar\alpha_0$. Event-shape variables, like thrust, broadening, $C$
parameter and jet mass (inspired by $\ele$ measurements), are well
suited to test the universality of this effective coupling. In this
type of analysis, the data are compared to a model prediction which
consists of a combination of NLO QCD calculations and the expectations
of the power-correction model: 
$F=F_{\rm perturbative}+F_{\rm power\ correction}$, where $F$ is an
event-shape mean or distribution. In the case of distributions, the
pertubative prediction is supplemented by matched resummed
calculations. Figure~\ref{fig3}(a) shows, as an example, the
measured thrust distribution in different regions of
$\q2$~\cite{desy-06-042} together with the predictions which have been
fitted to the data to extract $\bar\alpha_0$ and $\asz$. It is
possible to obtain a good description of the event-shape observables
in hadronic-induced reactions in the regions of phase space where the
predictions of the power-correction model are valid. The results
of the fits to extract $\bar\alpha_0$ and $\asz$ are shown, for all
event-shape observables measured, in
Fig.~\ref{fig3}(b)~\cite{epj:c46:343}. The extracted $\asz$ values are
consistent for all observables and with the world average (shaded
band). It is possible to extract a universal non-perturbative
parameter, $\bar\alpha_0=0.5\pm 10\%$, from all the measured
observables. This supports the concept of power corrections as an
appropiate alternative approach for the description of hadronisation
effects.

%\vspace{-0.5cm}
\subsection{Parton evolution at low $x$ and unintegrated PDFs}
One of the main channels of Higgs production at LHC is expected to be
$gg\rightarrow H$. Predictions for this process need information on
the parton evolution at low $x$ and on the unintegrated proton
PDFs. Forward-jet data at HERA are ideally suited to study these
effects.

At high scales ($Q$, $\etjet$), NLO calculations using the DGLAP
evolution equations provide a good description of the data (see
Section 3). Thus, measurements at HERA have provided accurate
determinations of the proton PDFs and precise determinations of
$\as$. The DGLAP evolution is equivalent to the exchange of a parton
cascade in which the exchanged partons are strongly ordered in
virtuality $\kt$. This approximation works well at high scales, as
mentioned above, but it is expected to break down at low $x$ since it
includes only resummation of leading logarithms in $\q2$ and neglects
contributions from $\log 1/x$, which are important when
$\log\q2\ll\log 1/x$. Approaches to parton evolution which deal with
low $x$ comprise: ($i$) BFKL evolution, which includes resummation of
$\log 1/x$ terms and exhibits no $\kt$ ordering; ($ii$) the CCFM
equations, which incorporate angular-ordered parton emission and 
are equivalent to BFKL for $x\rightarrow 0$ and to DGLAP at large $x$;
($iii$) the introduction of a second $\kt$-ordered parton cascade on the
photon side \`{a} la DGLAP, which is implemented by assigning a
partonic structure to the virtual photon, mimicks higher-order QCD
effects at low $x$. HERA is an ideal testbed for these theoretical
approaches. These tests are realised by restricting the jet data to
large jet pseudorapidity values (forward direction, proton side),
$x_{\rm jet}\equiv\etjet/E_p\gg x_{\rm Bj}$ (to suppress
quark-parton-model type events) and $Q\approx\etjet$ (to restrict
evolution in $\q2$).

Figure~\ref{fig4} shows the measured forward-jet
data~\cite{epj:c46:27} as a function of Bjorken $x$. The data
increases as $x_{\rm Bj}$ decreases. The data are compared to
different predictions. The NLO pQCD calculations,
based on DGLAP evolution equations, fail to describe the data at low
$x$. The predictions of the {\sc Cascade} Monte Carlo (MC), which is
an implementation of the CCFM equations, give an improved description
of the data at low $x$, but the description at high $x$ worsens. The
predictions of the CDM model, which contain no-$\kt$ ordering, and the
resolved-photon MC give a better description of the data, except at
very low $x$. 

Insight into low-$x$ dynamics can also be gained by studying the
azimuthal separation ($\Delta\phi$) between the two hardest jets. An
excess of events at small $\Delta\phi$ would signal a deviation from
DGLAP evolution. The fraction $S$ of dijet events with
$\Delta\phi<120$ has been measured~\cite{epj:c33:477} as a function of
Bjorken $x$ in different regions of $\q2$. The data are shown in
Fig.~\ref{fig5} compared with different predictions. The predictions
of the CDM model give a good description of the data at low $x$ and
low $\q2$, but are below the data for high $\q2$. The predictions of
the {\sc Cascade} MC using different sets of unintegrated PDFs are
also shown: the predictions display sensitivity to the unintegrated
gluon distributions. Therefore, these measurements can be used to
constrain the unintegrated PDFs.

%\vspace{-0.5cm}
\section{QCD studies in photoproduction}

%\vspace{-0.2cm}
\subsection{Test of color dynamics in dijet PHP}
Measurements of jet cross sections in PHP allow tests of color
dynamics. At HERA, quark and gluon exchange can be studied in dijet
PHP by separating resolved and direct processes using the fraction of
the photon energy invested in the production of the dijet system,
which is measured via the observable 
$\xo=\sum E_T^{{\rm jet}_i}e^{-\eta^{{\rm jet}_i}}/2yE_e$. For direct
events, $\xo\sim 1$ and for resolved events, $\xo<1$. Resolved
processes are dominated by gluon exchange (like dijets in $pp$
collisions) and the angular behaviour of the cross sections is 
$\propto (1-\cost)^{-2}$ for $\theta^*\rightarrow 0$ (Rutherford
scattering), where $\theta^*$ is the scattering angle in the dijet
centre-of-mass frame. Direct processes proceed via quark exchange
(similar to prompt-photon processes in $pp$ collisions) and the
angular behaviour of the cross section is $\propto (1-\cost)^{-1}$
for $\theta^*\rightarrow 0$. Therefore, the $\cos\theta^*$
distribution reflects the underlying parton dynamics since it has
sensitivity to the spin of the exchanged particle in two-body
processes.

The data sample for $\xo\!>\!(<)0.75$ is expected to be dominated by
direct (resolved) processes, as demonstrated by the $\xo$ distribution
for data and resolved, direct and resolved+direct MC events shown in
Fig.~\ref{fig6}(a)~\cite{pl:b384:401}. These samples were used to
compute the cross section as a function of $\cost$ (see
Fig.~\ref{fig6}(b))~\cite{pl:b384:401}. The cross sections increase as
$\cost$ increases, but the cross section for $\xo<0.75$ displays a
much more rapid increase for $\cost\rightarrow 1$ than the cross
section for $\xo>0.75$. The LO and NLO QCD calculations, which include
resolved (direct) processes with an angular behaviour 
$\propto (1-\cost)^{-2(-1)}$ give a very good description of the
data. This demonstrates that the underlying parton dynamics in PHP is
well understood.

%\vspace{-0.5cm}
\subsection{Inclusive-jet cross sections, $\as$ and the photon PDFs}
Measurements of jet cross sections in PHP allow tests of
pQCD. Inclusive-jet cross sections have been measured in PHP at
large $\etjet$ ($\etjet>17$~GeV~\cite{pl:b560:7} and
$\etjet>21$~GeV~\cite{epj:c29:497}) as functions of $\etjet$ and
$\etajet$ (see Figs.~\ref{fig7}(a) and (c), respectively). The
measured cross section as a function of $\etjet$ displays a steep
fall-off of several orders of magnitude within the measured range. The
measurements have been compared to NLO QCD calculations. The
predictions give a good description of the data. This proves the
validity of the description of the dynamics of inclusive-jet
production by pQCD at $\oalphas2$. The QCD prediction for the jet
cross section is directly proportional to $\as$. A value of $\asz$ has
been extracted from the measurement of $\set$:

\centerline{$\asmz{0.1224}{0.0001}{0.0019}{0.0022}{0.0042}{0.0054}.$}
The small experimental ($1.8\%$) and theoretical ($4.4\%$)
uncertainties make this determination also one of the most precise at
HERA. This value and the ones quoted in Section 3.1 are in agreement
with the world~\cite{bethke} and HERA~\cite{herave} averages. The
energy-scale dependence of $\as$ has been tested from the measurements
of the inclusive-jet cross section as a function of $\etjet$. The
results, shown in Fig.~\ref{fig7}(b), are in very good agreement with
the running of $\as$ as predicted by pQCD over a wide range in
$\etjet$.

Measurements of jet cross sections in PHP also allow tests of the
photon PDFs. The structure of the photon can be investigated at HERA
by measuring jet cross sections most sensitive to the photon PDFs,
namely $\seta$ (see Fig.~\ref{fig7}(c))~\cite{epj:c29:497} or $\sxo$,
and comparing the measurements to predictions based on different
parametrisations of the PDFs. The measurements can then be used to
discriminate among different parametrisations or used in a global fit
to constrain them, as it was recently done for the proton PDFs (see
Section~5).

%\vspace{-0.5cm}
\subsection{Multijet cross sections and multiparton interactions}
Measurements of jet cross sections in PHP allow tests of hard multijet
production and multiparton interactions (MPIs), which are expected to
be copious at LHC energies. Multijet production
is directly sensitive to higher orders since e.g. the three-jet cross
section is proportional to $\as^2$ at lowest order. Such measurements
allow tests of the parton showers in the MC models. Their sensitivity
to MPIs allows tests and tuning of the models. Figure~\ref{fig8} shows
the three-jet cross section for $\m3j>25$~GeV as functions of $\xo$
and $y$~\cite{zeus-prel-06-009}. The data are compared to the
predictions of the {\sc Pythia} and {\sc Herwig} MC models which
describe the shape of the $y$ distribution but fail to describe the
shape of the measured $\xo$ distribution. The inclusion of MPIs in
{\sc Pythia}, tuned to generic collider data, results in a failure to
describe both data distributions. In the case of {\sc Herwig}, the MPI
model was tuned to the $\xo$ distribution and gives a good description
of the data, but the description of $y$ gets spoiled. Therefore, these
very precise data are an ideal ground for tuning and testing the 
parton-shower and MPI models in a clean hadronic-induced enviroment. 

%\vspace{-0.5cm}
\section{Jet cross sections and the proton PDFs}
Measurements of jet cross sections in NC DIS and PHP are directly
sensitive to the proton PDFs and provide a useful constrain on the
gluon density, especially at mid- to high-$x$ values, which are most
relevant at LHC energies. This is achieved in the case of PHP by
restricting the measurements to regions of phase space less sensitive
to the photon PDFs, i.e. $\xo>0.75$. Recently, very precise
measurements of inclusive-jet cross sections in NC DIS as a function
of $\etjb$ in different regions of $\q2$ (see
Fig.~\ref{fig9}(a))~\cite{pl:b547:164} and dijet cross sections in PHP
as a function of $\etjet$ in different $\etajet$ regions for
$\xo>0.75$ (see Fig.~\ref{fig9}(b))~\cite{epj:c23:4} have been
incorporated in a rigorous way in a QCD fit to determine the proton
PDFs~\cite{epj:c42:1}. The valence, sea and gluon distributions as a
function of $x$ for $\q2=10$~\g2\ are shown in
Fig.~\ref{fig10}(a). The Incorporation of jet cross sections in the fit
resulted in an improvement on the determination of the gluon density
in the proton: as shown in Fig.~\ref{fig10}(b), the uncertainty in the
gluon density decreases for mid- to high-$x$ by up to a factor of two.

%\vspace{-0.5cm}
\section{Conclusions}
HERA has become an unique QCD-testing machine, very useful for
understanding multijet production in clean hadronic-induced
reactions. Considerable progress in understanding and reducing 
uncertainties has led to very precise determinations of $\as$ (see
Fig.~\ref{fig11}(a)). The dominant uncertainty in these measurements
comes from the thoretical side thus, improved QCD calculations are
needed for better accuracy. The running of $\as$ has been observed
from HERA jet data alone (see Fig.~\ref{fig11}(b)). New HERA I results
are still coming; new determinations of $\asz$ are shown in
Fig.~\ref{fig11}(c) and more than 400~\pb1\ of integrated luminosity of
$e^{\pm}p$ data with polarised lepton beams are being analysed.
Therefore, there is a wealth of hadronic-type-reaction data already
available to test new MC models, MC@NLO and
next-to-next-to-leading-order pQCD calculations if they are made
available for $ep$ collisions.

%\vspace{-0.5cm}

\end{document}